\documentclass[prd,a4paper,twocolumn]{revtex4}
\usepackage{tikz}
\usepackage{graphicx}
\usepackage[]{subfigure}
\usepackage{epsfig}
\usepackage{color}
\definecolor{Blue}{rgb}{0.1,0.1,1.0} 
\definecolor{Magenta}{rgb}{1.0,0.1,0.5}

\newcommand{\nc}{\newcommand}

\nc{\be}[1]{\begin{equation}\mbox{$\label{#1}$}}
\nc{\bea}[1]{\begin{eqnarray} \mbox{$\label{#1}$}}
\nc{\Section}[2]{\section{#2}\label{#1}}
\nc{\Bibitem}[1]{\bibitem{#1}}
\nc{\Label}[1]{\label{#1}}

\nc{\eea}{\end{eqnarray}}
\nc{\ee}{\end{equation}}

\newcommand{\ket}[1]{| #1 \rangle}

\nc{\bdm}{\begin{displaymath}}
\nc{\edm}{\end{displaymath}}
\nc{\dpsty}{\displaystyle}
\nc{\bc}{\begin{center}}
\nc{\ec}{\end{center}}
\nc{\ba}{\begin{array}}
\nc{\ea}{\end{array}}
\nc{\bab}{\begin{abstract}}
\nc{\eab}{\end{abstract}}
\nc{\btab}{\begin{tabular}}
\nc{\etab}{\end{tabular}}
\nc{\bit}{\begin{itemize}}
\nc{\eit}{\end{itemize}}
\nc{\ben}{\begin{enumerate}}
\nc{\een}{\end{enumerate}}
\nc{\bfig}{\begin{figure}}
\nc{\efig}{\end{figure}}

\nc{\arreq}{&\!=\!&}
\nc{\arrmi}{&\!-\!&}
\nc{\arrpl}{&\!+\!&}
\nc{\arrap}{&\!\!\!\approx\!\!\!&}
\nc{\non}{\nonumber}
\nc{\align}{\!\!\!\!\!\!\!\!&&}

\def\lsim{\; \raise0.3ex\hbox{$<$\kern-0.75em
      \raise-1.1ex\hbox{$\sim$}}\; }
\def\gsim{\; \raise0.3ex\hbox{$>$\kern-0.75em
      \raise-1.1ex\hbox{$\sim$}}\; }

\nc{\DOT}{\hspace{-0.08in}{\bf .}\hspace{0.1in}}
\nc{\Laada}{\hbox {$\sqcap$ \kern -1em $\sqcup$}}
\nc\loota{{\scriptstyle\sqcap\kern-0.55em\hbox{$\scriptstyle\sqcup$}}}
\nc\Loota{{\sqcap\kern-0.65em\hbox{$\sqcup$}}}
\nc\laada{\Loota}
\nc{\qed}{\hskip 3em \hbox{\BOX} \vskip 2ex}

\nc{\real}{{\rm I \! R}}
\nc{\Z}{{\sf Z \!\!\! Z}}
\nc{\complex}{{\rm C\!\!\! {\sf I}\,\,}}
\def\bigid{\leavevmode\hbox{\small1\kern-3.8pt\normalsize1}}
\def\id{\leavevmode\hbox{\small1\kern-3.3pt\normalsize1}}
\nc{\slask}{\!\!\!/}
\nc{\bis}{{\prime\prime}}
\nc{\pa}{\partial}
\nc{\na}{\nabla}
\nc{\ra}{\rangle}
\nc{\la}{\langle}
\nc{\goto}{\rightarrow}
\nc{\swap}{\leftrightarrow}

\nc{\EE}[1]{ \mbox{$\cdot10^{#1}$} }
\nc{\abs}[1]{\left|#1\right|}
\nc{\at}[2]{\left.#1\right|_{#2}}
\nc{\norm}[1]{\|#1\|}
\nc{\abscut}[2]{\Abs{#1}_{\scriptscriptstyle#2}}
\nc{\vek}[1]{{\rm\bf #1}}
\nc{\integral}[2]{\int\limits_{#1}^{#2}}
\nc{\inv}[1]{\frac{1}{#1}}
\nc{\dd}[2]{{{\partial #1}\over{\partial #2}}}
\nc{\ddd}[2]{{{{\partial}^2 #1}\over{\partial {#2}^2}}}
\nc{\dddd}[3]{{{{\partial}^2 #1}\over
    {\partial #2 \partial #3}}}
\nc{\dder}[2]{{{d #1}\over{d #2}}}
\nc{\ddder}[2]{{{d^2 #1}\over{d {#2}^2}}}
\nc{\dddder}[3]{{d^2 #1}\over
    {d #2 d #3}}
\nc{\dx}[1]{d\,^{#1}x}
\nc{\dy}[1]{d\,^{#1}y}
\nc{\dz}[1]{d\,^{#1}z}
\nc{\dl}[1]{\frac{d\,^{#1}l}{(2\pi)^{#1}}}
\nc{\dk}[1]{\frac{d\,^{#1}k}{(2\pi)^{#1}}}
\nc{\dq}[1]{\frac{d\,^{#1}q}{(2\pi)^{#1}}}

\nc{\bfT}{{\bf T }}

\nc{\cA}{{\cal A}}
\nc{\cB}{{\cal B}}
\nc{\cD}{{\cal D}}
\nc{\cE}{{\cal E}}
\nc{\cG}{{\cal G}}
\nc{\cH}{{\cal H}}
\nc{\cL}{{\cal L}}
\nc{\cO}{{\cal O}}
\nc{\cT}{{\cal T}}
\nc{\cN}{{\cal N}}
\nc{\cR}{{\cal R}}

\nc{\rvac}[1]{|{\cal O}#1\rangle}
\nc{\lvac}[1]{\langle{\cal O}#1|}
\nc{\rvacb}[1]{|{\cal O}_\beta #1\rangle}
\nc{\lvacb}[1]{\langle{\cal O}_\beta #1 |}
\nc{\bb}{\bar{\beta}}
\nc{\bt}{\tilde{\beta}}
\nc{\ctH}{\tilde{\cal H}}
\nc{\chH}{\hat{\cal H}}
\nc{\al}{\alpha}
\nc{\g}{\gamma}
\nc{\Del}{\Delta}
\nc{\e}{\textrm{e}}
\nc{\eps}{\epsilon}
\nc{\lam}{\lambda}
\nc{\Om}{\Omega}
\nc{\ve}{\varepsilon}
\nc{\mn}{{\mu\nu}}
\nc{\vp}{\varphi}

\nc{\rf}[1]{(\ref{#1})}
\nc{\nn}{\nonumber \\*}
\nc{\bfB}{\bf{B}}
\nc{\bfv}{\bf{v}}
\nc{\bfx}{\bf{x}}
\nc{\bfy}{\bf{y}}
\nc{\vx}{\vec{x}}
\nc{\vy}{\vec{y}}
\nc{\oB}{\overline{B}}
\nc{\oI}{\overline{I}}
\nc{\oR}{\overline{R}}
\nc{\rar}{\rightarrow}
\nc{\ti}{\times}
\nc{\slsh}{\hskip-5pt/}
\nc{\sm}{Standard~Model~}
\nc{\MP}{M_{\rm Pl}}
\nc{\mpl}{M_{\rm Pl}}
\nc{\tp}{t_{\rm Pl}}

\nc{\pmin}{p_{\rm min}}
\nc{\pmax}{p_{\rm max}}
\nc{\fo}{f_0}
\nc{\foi}{f_{0,i}\,}
\nc{\fop}{f_0^P}
\nc{\fou}{f_0^U}

\nc{\eff}{{\rm eff}}
\nc{\MT}{M_{\rm T}}
\nc{\ML}{M_{\rm L}}
\nc{\kk}{\vek{k}}
\nc{\pp}{{\rm p}}
\nc{\pt}{\partial_t}
\nc{\half}{{1\over 2}}
\nc{\w}{\omega}
\nc{\uhat}{\hat{U}_\w}

\nc{\etal}{\mbox{\it et al.}}
\nc{\ie}{{\it i.e. }}
\nc{\eg}{{\it e.g. }}
\nc{\trh}{T_{\rm RH}}
\nc{\ad}{{a'\over a}}
\nc{\bd}{{b'\over b}}
\nc{\Rd}{{R'\over R}}
\nc{\diag}{{\textrm{diag}}}
\nc{\mato}[1]{\tilde{#1}}
\nc{\sinn}{\textrm{sinn}}
\nc{\sech}{\textrm{sech}}
\nc{\I}{\textrm{I}}
\nc{\II}{\textrm{II}}
\nc{\III}{\textrm{III}}
\nc{\vev}[1]{\langle #1 \rangle}
\nc{\hyp}{\,\; F_{1{\hskip -16pt}2}{\hskip 11pt}}
\nc{\brhom}{\overline{\rho}_M}
\nc{\brho}{\overline{\rho}}
\nc{\rhob}{\overline{\rho}}
\nc{\Pb}{\overline{P}}
\nc{\bH}{\overline{H}}
\nc{\ep}{{1+4\eps}}

\nc{\deriv}[2]{ 
\frac{\mathrm{d}#1}{\mathrm{d}#2}
}
\nc{\Mnu}{M_\nu}
\nc{\bee}{\begin{equation}}
\nc{\ene}{\end{equation}}
\nc{\hdp}{\sigma_8 (\Omega_{\rm m}/0.3)^{0.37}}
\nc{\avis}{\alpha_{vis}}
\nc{\cvis}{c^2_{vis}}
\nc{\clam}{c^2_{lam}}

\def\smiley{\hbox{\large$\bigcirc$\hspace{-.80em}%
\raise.2ex\hbox{$\cdot\cdot$}\kern-.61em    %--- .56
\lower.2ex\hbox{\scriptsize$\smile$}}\ }

\def\frowney{\hbox{\large$\bigcirc$\hspace{-.80em}%
\raise.2ex\hbox{$\cdot\cdot$}\kern-.635em
\lower.2ex\hbox{\scriptsize$\frown$}}\ }

\begin{document}

\title{Detection of transplanckian effects in the cosmic microwave background}
\author{Nicolaas E. Groeneboom}
\email{nicolaag@astro.uio.no}
\affiliation{Institute of Theoretical Astrophysics, University of Oslo, Box 1029, 0315 Oslo, Norway}

\author{\O ystein Elgar\o y}
\email{oystein.elgaroy@astro.uio.no}
\affiliation{Institute of Theoretical Astrophysics, University of Oslo, Box 1029, 0315 Oslo, Norway}

\date{\today}

\begin{abstract} 
Quantum gravity effects are expected modify the primordial density fluctuations produced during inflation and leave their imprint on the cosmic microwave background observed today. 
We present a new analysis discussing whether these effects are detectable, considering both currently available data and 
simulated results from an optimal CMB experiment. We find that the WMAP data show no evidence for 
the particular signature considered in this work, but give an upper bound on the parameters of the model. 
However, a hypothetical experiment shows that with proper data, the transplanckian effects
should be detectable through alternate sampling methods. This fuzzy conclusion 
is a result of the nature of the oscillations, since they give rise to a likelihood hyper-surface riddled with local maxima.  
A simple Bayesian analysis shows no significant evidence
for the simulated data to prefer a transplanckian model. 
Conventional Markov Chain Monte Carlo (MCMC) methods are not suitable for exploring this complicated landscape, but alternative methods 
are required to solve the problem. This however requires extremely high-precision data.
%We conclude by explaining why it is very unlikely that 
%transplanckian effects can be detected convincingly by CMB data alone in the near future, 
%even for optimistic assumptions about the energy scale of quantum gravity.  
\end{abstract}

\maketitle

\section{Introduction}   
Quantum fluctuations in the scalar field responsible for driving inflation can give rise to a power spectrum of primordial density perturbations consistent with what is required to seed the large-scale structure (LSS) and the temperature anisotropies in the cosmic microwave background (CMB) radiation.  This fact is probably one of the main reasons for inflation having become a part of the concordance model of cosmology today.  The quantum fluctuations are imprinted during the 
early inflationary epoch, and become classical fluctuations in the gravitational 
potential as the fluctuations leave the causal horizon.  The standard way 
of calculating the power spectrum of the density fluctuations make use 
of the fact that space-time becomes Minkowskian in the 
distant past.  However, it has been argued \cite{ martin-2001-63, easther-2001-64,ulf1, ulf2, jerome1, jerome2, jerome3, jerome4, bergstrom} 
that since a given length scale observable in the universe today shrinks and 
becomes smaller than the Planck length if followed sufficiently far back in time, effects of quantum gravity will at some point play a role in setting up 
the perturbations, and may potentially leave an observable signature. 
The primordial density fluctuations in the early universe may have  
small oscillations superimposed as a result of quantum gravitational effects.
These effects are due to non-negligible curvature imposed by high energy densities in de Sitter space during the early stages of inflation.
Several papers have suggested a generic shape of these effects \cite{ulf1, ulf2, bergstrom} in the form of small oscillations superimposed on the standard, 
nearly scale-invariant primordial power spectrum of density fluctuations. 
A crucial questions is, of course, whether such a signature is observable 
in the universe today.  This question has been investigated in several  
papers \cite{jerome1, jerome2, jerome3, jerome4, elgen, ulf1, ulf2}.  The conclusion in \cite{elgen} was pessimistic: the so-called 
transplanckian effects were found to be unobservable in practice.  
However, this conclusion has been called into question in \cite{jerome4}.  
One of the problems pointed out in that work is that the oscillations 
require more numerical accuracy than what is commonly employed in 
codes like CMBFAST \cite{Seljak:1996is} and CAMB \cite{Lewis:1999bs}.  

In this paper, we revisit the question of the observability of transplanckian 
effects in the CMB, using a more accurate numerical treatment than in 
\cite{elgen}.  Both present data and a hypothetical optimal CMB experiment 
are considered.  Furthermore, we mention alternative sampling algorithms to 
overcome the numerical problems encountered in exploring what turns out 
to be a particularly nasty likelihood hyper-surface. Section II gives 
a brief description of the particular model we consider and the data sets 
we use. In section III we discuss some technical 
details in the numerical computations undertaken.  Section IV then 
contains results obtained with currently available data sets, and in 
section V we investigate the possibilities with an ideal CMB data set. The
exact likelihood function is described in section VI, while a Bayesian
evidence analysis is performed in section VII. 
A discussion of the results and a suggestion for an improved approach 
to the search for transplanckian effects can be found in section VIII, and we 
conclude in section IX. 

\section{Background}
\label{background}
Neglecting the perturbations in the metric during inflation, it is viable to assume that the inflaton scalar field can be expressed as:
\begin{equation}
\phi(\textbf x,t) = \underbrace{\phi_0(t)}_{\text{classical evolution}} + \underbrace{\delta \phi(\textbf x,t)}_{\text{fluctuations}} 
\end{equation}
The fluctuations in a scalar field $\delta \phi$ in a Friedmann-Lema\^itre-Robertson-Walker background satisfy the equation of motion (EOM):
\begin{equation}
\label{eompert}
 \ddot {\delta \phi} + 3H\dot {\delta \phi} - 
\frac{1}{a}(\nabla^2)\delta \phi +m^2\delta \phi  = 0 
\end{equation}
where the dot denotes the derivative with respect to time and $m^2 \propto 
d^2 V/d\phi^2$, where $V(\phi)$ is the inflaton potential. 
Performing a Fourier expansion of the fluctuations: 
\begin{equation}
\delta \phi( x,t) = \frac{1}{\sqrt N}\sum_k u_k(t)e^{ikx} 
\end{equation}
where $N$ is a normalization constant, 
equation (\ref{eompert}) becomes:
\begin{equation}
  \ddot u_k + 3H \dot u_k + (\frac{k^2}{a^2}  + m^2)u_k = 0.
\end{equation}
Assuming de Sitter space, 
this equation can be rewritten using a change of variable ($u_k \equiv \varphi/a$) and switching to conformal time ($d\eta = dt/a$).
This removes the friction term, 
and the final EOM becomes:
\begin{equation}
\varphi_k'' +  \Big(k^2 - \frac{2}{\eta^2}\Big)\varphi_k  = 0
\end{equation}
where $'$ denotes derivative with respect to conformal time $\eta$. A detailed derivation of this result can be obtained in \cite{dodelson:book}.
The general solution for $\varphi_k$ is:
\begin{equation}
\varphi_k = 
\frac{A_k}{\sqrt{2 k}}e^{-ik\eta}\Big(1-\frac{i}{k\eta}\Big) + 
\frac{B_k}{\sqrt{2 k}}e^{ik\eta}\Big(1+\frac{i}{k\eta}\Big)  
\end{equation}
where $A_k$ and $B_k$ are Bogoliubov coefficients satisfying $|A_k|^2 -|B_k|^2 = 1$, $\eta$ is the conformal time and $k$ is the wave mode. 
The usual choice of vacuum is the Bunch-Davies vacuum, which corresponds to $A_k =1$ and $B_k=0$. Space-time becomes Minkowskian
in the infinite past ($\eta \to - \infty)$, where it is feasible to impose initial conditions on the field equations.

In
\cite{ulf2}, different possibilities for suitable vacua are discussed before focusing on a specific choice, the adiabatic
vacuum, for which $a_k(\eta_0) \ket{0, \eta_0} = 0$. This vacuum (for a given $\eta_0$) is a typical representative of vacua, and corresponds
to a minimum uncertainty between field and its conjugate momentum \cite{polarski}. When imposing
initial conditions, Bogoliubov-coefficients are constrained by the relation \cite{ulf2}:
\begin{equation}
B_k = \frac{ie^{-2ik\eta_0}}{2k\eta_0 +i} A_k.
\end{equation}
Note that the Bunch-Davies vacuum is restored in the infinite past: When $\eta_0 \to -\infty$, $A_k = 1$ and $B_k = 0$. A discussion of the
implications of the choice of vacuum can be found in \cite{Bozza:hep-th0302184, campo-2007}.

When choosing this zeroth order adiabatic vacuum \cite{ulf1}, 
the primordial curvature spectrum can be expressed as \cite{ulf2}:
\begin{equation}
 P(k) = \Big( \frac{H}{\dot \phi} \Big)^2 
\Big(\frac{H}{2\pi}\Big)^2 \Big(1 - \frac{H}{\Lambda} \sin \big(\frac{2\Lambda}{H}\big)\Big)
\label{mod}
\end{equation}
where $\Lambda$ is the (Planck) energy cutoff scale and $H$ the Hubble parameter. Note that the size
of the corrections is \textit{linear}  \cite{kaloper-2002-66} in $H/\Lambda$. In \cite{kaloper}, it is  argued that equation (\ref{mod}) represents
a generic effect, regardless of the details in the (undetermined) transplanckian physics. In \cite{ulf2}, equation
(\ref{mod}) is further parametrized as
\begin{equation}
P(\epsilon, \xi, k) = P_0(k)\Big(1-\xi \Big(\frac{k}{k_n}\Big)^{-\epsilon} 
\sin \Big[ \frac{2}{\xi}\Big(\frac{k}{k_n}\Big)^{\epsilon}\Big] \Big)
\label{finalmp}
\end{equation}
with variables explained in table \ref{table1}.
\begin{table}[h!t!b!]
\center
\begin{tabular}{ r r r }
 \hline
 \hline
 \small
  \small \textbf{Parameter} & \textbf{Value}  & \textbf{Description}\\
 \hline    
  \small 
  $\epsilon$ & $\epsilon \ll 1$ & The slow roll parameter \\
  $\gamma$ & $\Lambda / M_{pl}$ & The transplanckian scale \\
  $M_{pl} $ & $1/\sqrt{8 \pi G}$ & The reduced Planck mass \\
  $\xi$ & $4\cdot 10^{-4} \frac{\sqrt \epsilon}{\gamma}$  & A parametrization of $H/\Lambda$  \\
  $k_n$ & $k_n$ & The largest measurable scale\\
  $P_0(k)$ &  $\propto k^{n_s -1}$& The standard power spectrum\\ 
 \hline
\end{tabular}
 \caption{Explaining equation (\ref{finalmp})}
 \label{table1}
\end{table}
\normalsize
Notice how the standard inflationary power spectrum $P_0$ is restored as the cutoff $\xi \propto \frac{1}{\Lambda} \to 0$. It is also interesting
to note that for low cutoff values $\xi \propto \frac{1}{\Lambda} \to \infty$, the amplitude of the modulation converges to $2$ (as $\lim_{x \to \infty} x \sin (2/x) = 2$). The
amplitude of the oscillations never reaches values greater than $2$, and only severely impacts $P_0$ for large values of $\xi$. Physically, if the cutoff 
$\xi \propto \frac{1}{\Lambda}$ was enforced at low values (large $\xi$), it should easily be detectable in the the cosmic microwave background,
as will be presented in the following section. This is, however, not the case, and 
we know that the transplanckian physics leave imprints at $\xi \ll 1$, corresponding to a cutoff at high energies. The amplitude of the oscillations 
is therefore expected to be very small.

\begin{figure}[h!t!b!]
  \begin{center}
    \includegraphics[width=70mm]{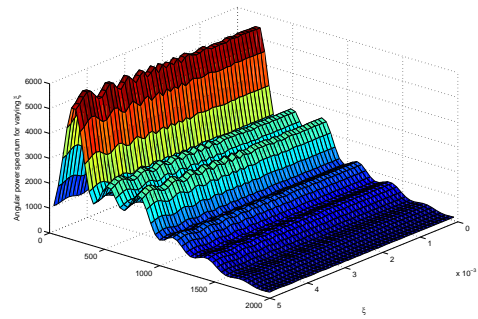}
  \end{center}
  \caption{The full power spectrum $C_\ell$ for varying $\xi$.}
\label{varyingxi1}
\end{figure}

\begin{figure}[h!t!b!]
  \begin{center}
    \includegraphics[width=70mm]{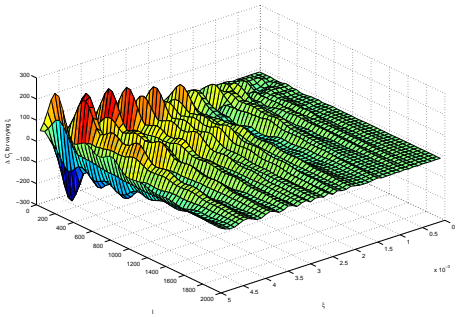}
  \end{center}
  \caption{The transplanckian modulations $\Delta C_\ell$ for varying $\xi$.}
\label{varyingxi2}
\end{figure}

When evolving the modulated primordial power spectrum through CAMB,  
the transplanckian imprints for varying $\xi$ in the full power spectrum can be seen in 
figure \ref{varyingxi1}. The oscillations singled out for varying $\xi$ are presented in figure \ref{varyingxi2}.  
%
%In \cite{jerome1}, it is argued that the amplitude of these oscillations should be decoupled from
%equation (\ref{finalmp}) and be treated as a freely varying parameter.  
%This argument seems to be contrary to the consensus found in the literature that the transplanckian effects are at most of order $H/\Lambda$.  
%If the amplitude is left as a free parameter in fits to cosmological data, 
%one effectively looks for a more general oscillatory feature.  The connection 
%to genuine transplanckian features is then lost.  

\subsection{A brief summary of recent results}
Reference \cite{ulf1} concludes that effects of transplanckian
physics are possibly within the reach of cosmological observations.  
Equation (\ref{finalmp}) describes a
a generic expression for how transplanckian effects would modify the primordial power spectrum.
As seen in the previous section, these oscillations are caused by a nontrivial vacuum for the inflaton
field \cite{ulf1, ulf2}. As the oscillations are expected to contribute to the energy density, this 
could change the way the universe expands, and in a worst-case scenario, the inflationary phase could be destroyed. 
The authors of \cite{ulf2} investigate this possibility, and conclude that the the back reaction is under control and fully consistent
with inflation, with a slow roll found to be completely dominated by the vacuum energy given the parameters suggested in \cite{jerome2}. 

\subsubsection{WMAP data and transplanckian effects}
A previous data analysis \cite{brandenberger} concluded that no significant signals from transplanckian effects are visible 
in the CMB. Another analysis \cite{jerome2}
claimed that there are some weak hints in the current data, and these indications have become slightly stronger
with the WMAP3 data compared to earlier claims by the same authors \cite{jerome1, jerome2, jerome3, jerome4}. The parameters
implied by the data suggests oscillations in amplitude
that are periodic in the logarithm of the scale of the CMB fluctuations, just as predicted from transplanckian physics.\\

In \cite{jerome1}, the authors discuss the so-called cosmic variance outliers, i.e.
points which lie outside the $1\sigma$ cosmic variance error. These outliers are considered interesting  as the probability of
their presence is very small \cite{spergel}. The authors mention that it has been envisaged that the outliers could be
a signature of new physics, even though the cosmic variance could be responsible for their presence. The conclusion of \cite{jerome1}
is that there exist statistical  justification for a presence of oscillations in the power spectrum. It should however
be noted that the authors of \cite{jerome1} treat the \textit{amplitude} of the oscillations as a free parameter, which
ensures the fitting of the cosmic variance outliers with the transplanckian oscillations. \\
%As explained in the introduction, we are reluctant to classify these modulations as transplanckian.\\

Another paper \cite{elgen} was less optimistic, and concluded that it is unlikely that a transplanckian signature of this type can be detected in CMB
and large-scale structure data.  However, \cite{elgen} left some 
technical issues open, and we find it worthwhile to revisit this problem.
We show that the current WMAP3 data gives weak constraints on the parameters of the transplanckian model  
using conventional MCMC methods. 

\subsubsection{Simulated data and transplanckian effects}
The conclusion of \cite{elgen} was that CMB and LSS data are in principle sensitive to transplanckian modulations in the primordial power
spectrum, but that is practically impossible to make a positive detection even in future high-precision data. This has to do
with the \textit{nature} of the oscillations, as the value of the likelihood function is extremely sensitive to $\epsilon$ and $\xi$.
But what ``extremely sensitive'' means was not explained in \cite{elgen}. We will show explicit examples of 
how this sensitive likelihood function contemplates trouble, and why this renders the underlying MCMC method in CosmoMC of little use in estimating
the transplanckian parameters.

\section{Numerical details}
\begin{figure}[h!t!b!]
  \begin{center}
    \includegraphics[width=70mm]{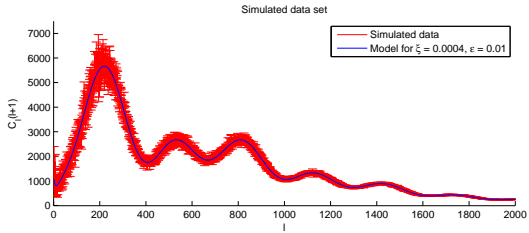}
  \end{center}
  \caption{Simulated $TT$ data based on $\xi = 0.0004$ and $\epsilon=0.01$.}
\label{simualteddata}
\end{figure}

CAMB \cite{Lewis:1999bs} was modified to include the primordial oscillations described in equation (\ref{finalmp}), adding the parameters $\xi$ and $\epsilon$.
Based on a power spectrum generated using the modified version of CAMB, 
a $TT$ data set was simulated by drawing from a $\chi^2$-distribution with $2\ell +1$ degrees of freedom. The data set and model 
are depicted in figure (\ref{simualteddata}), and the choice of values for $\xi = 0.0004$ and $\epsilon = 0.01$ 
are motivated by the Horava-Witten model \cite{hw1} of transplanckian physics \cite{ulf2}. In this model, unification
occurs at the scale where a fifth dimension becomes visible \cite{kaloper-2002-66}, resulting in quite optimistic estimates of the energy scale
where quantum gravity effects are manifested. However, the oscillations for this low value of $\xi$ corresponds to very small 
superimposed oscillations in the primordial power spectrum, as seen from figure \ref{varyingxi1} and \ref{varyingxi2}. 
The remaining cosmological parameters used for creating the data set (eg $\Omega_b, \tau, n_s$) were chosen as 
$\Lambda CDM$ best-fit values. We continued by creating a simulated $TE, EE$ and $BB$ data set using the same methods described here.

The MCMC software package CosmoMC \cite{lewis:2002} was altered to include the  parameters $\xi$ and $\epsilon$. 
Based on the likelihood for ideal noiseless simulated data described in \cite{verde2}, 
the WMAP-likelihood code of CosmoMC was adjusted to utilize the ideal data instead of the WMAP 3-year data. The TT-only likelihood function is given by 
\begin{equation}
\label{loglik}
 -2 \log \mathcal L(x_i; p_i) = \sum_\ell (2\ell +1)
\Big[
\ln \big(\frac{C_\ell}{\hat C_\ell}\big) + \frac{\hat C_\ell}{C_\ell} -1 
\Big].
\label{likeli1}
\end{equation}
where $C_\ell$ are the theoretical TT power spectra and $\hat C_\ell$ TT data sets.
The same method for creating and using ideal data with
 CosmoMC was employed in \cite{Mota:2007sz}. The errors in this ideal data set are only limited by cosmic variance, so no experiment can give any better estimate of the power spectrum. Thus, if transplanckian
effects are not detectable in the simulated data, they cannot be detectable with any  future high-precision data. We continued
by employing the full simulated $TT, TE, EE$ and $BB$ data set using the likelihood presented in \cite{gledesdreper}:
\begin{equation}
  -2 \ln \mathcal L = 
\sum_\ell (2\ell +1) \Big[
 \ln \big(\frac{C_\ell^{BB}}{\hat C_\ell^{BB}}\Big)
+\ln \big(\frac{C_\ell^{TT}C_\ell^{EE} - (C_\ell^{TE})^2}{\hat C_\ell^{TT}\hat C_\ell^{EE} - (\hat C_\ell^{TE})^2}\Big)
\label{likeli2}
\end{equation}
\begin{equation}
+\frac{\hat C_\ell^{TT}C_\ell^{EE} + C_\ell^{TT}\hat C_\ell^{EE} - 2C_\ell^{TE}\hat C_\ell^{TE}}
{C_\ell^{TT}C_\ell^{EE} - (C_\ell^{TE})^2}
\end{equation}
\begin{equation}
+\frac{\hat C_\ell^{BB}}{C_\ell^{BB}}
-3
\Big]
\end{equation}

\subsection{Increasing accuracy}

%\begin{figure}[h!t!b!]
%  \begin{center}
%       \includegraphics[width=50mm]{figures/mp_example/mp_difference_only1.eps}
%  \end{center}
%  \caption{Boosting accuracy: increasing accuracy\_boost.}
% \label{figacc1}
%\end{figure}

%\begin{figure}[h!t!b!]
%  \begin{center}
%       \includegraphics[width=50mm]{figures/mp_example/mp_difference_only2.eps}
%  \end{center}
%  \caption{Boosting accuracy: increasing l\_accuracy\_boost.}
% \label{figacc2}
%\end{figure}
%\subsubsection{l\_sample\_boost}
%\begin{figure}[h!t!b!]
%  \begin{center}
%       \includegraphics[width=50mm]{figures/mp_example/mp_difference_only3.eps}
%  \end{center}
%  \caption{Boosting accuracy: increasing l\_sample\_boost.}
% \label{figacc3}
%\end{figure}

\begin{figure}[h!t!b!]
  \begin{center}
       \includegraphics[width=80mm]{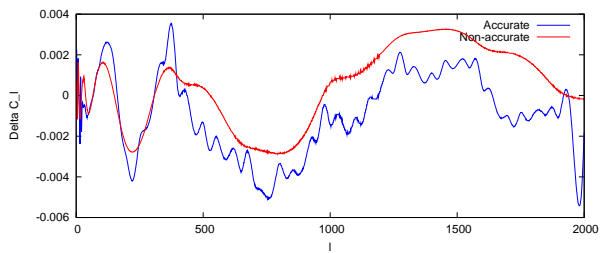}
  \end{center}
  \caption{The ratio of modulated power spectrum with low and high accuracy, all 3 parameters boosted.}
 \label{figaccuracy}
\end{figure}
Previous papers \cite{jerome3, jerome4}  have pointed out that it is necessary to increase the accuracy of the calculated power spectrum used for
estimating the transplanckian parameters. The results in \cite{elgen} might therefore have been partly due to the inaccuracy in the version of CMBFAST \cite{Seljak:1996is} used in this work. 
The accuracy of the numerical calculations was boosted, which increased run time by a tenfold. 
The results can be seen in figure \ref{figaccuracy}. However, the accuracy boosting had little or no effect on the WMAP parameter estimation,
as will be explained in the following section.

\section{Results from WMAP3 data}
\begin{figure}[h!t!b!]
  \begin{center}
    \includegraphics[width=40mm]{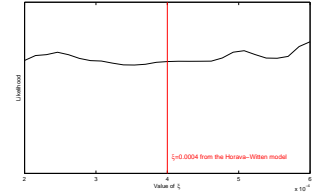}
  \end{center}
  \caption{Marginalized posterior for $\xi$ over a small interval surrounding the original input parameter $\xi=0.0004$ from the Horava-Witten model 
(red line).}
  \label{wmap2}
\end{figure}

\begin{figure}[h!t!b!]
  \begin{center}
    \includegraphics[width=40mm]{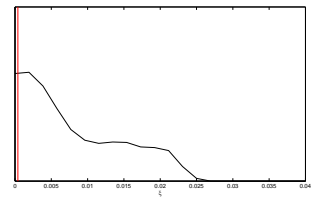}
  \end{center}
  \caption{Marginalized posterior of $\xi$ over a larger interval. The red line corresponds to $\xi=0.0004$ from the Horava-Witten model. Notice
how it is possible to estimate an upper bound: $\xi_{max} \sim 0.02$. Above this threshold, there are occasional ``bumps'' in the posterior
where the oscillation reaches zero ($\sin x = 0$, see equation \ref{finalmp}).}
  \label{wmapxi}
\end{figure}

\begin{figure}[h!t!b!]
  \begin{center}
    \includegraphics[width=40mm]{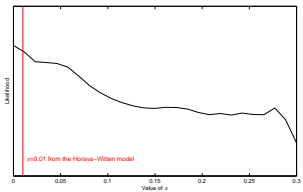}
  \end{center}
  \caption{Marginalized posterior for $\epsilon$. The red line corresponds to $\epsilon=0.01$ from the Horava-Witten model.}
  \label{wmap3}
\end{figure}

For small values of the transplanckian values, the constraints on both $\xi$ and $\epsilon$ from the WMAP3 data turned out rather poor when running CosmoMC with all standard cosmological parameters free. 
This 
can be seen in figure  \ref{wmap2} and \ref{wmap3}, consistent with the results in \cite{elgen}. For larger values of the transplanckian parameter $\xi$,
the increasing amplitude of the primordial power spectrum 
effectively constraints $\xi$ to the interval $[0,0.02]$ (recall from section \ref{background} that for large $\xi$, the amplitude converges to $2$). This can be seen in figure \ref{wmapxi}. This corresponds to a cutoff $\Lambda \sim
0.002 M_{pl} \sim 10^{16}GeV$. Everything below this scale ($\xi > 0.02$) should therefore be ruled out by WMAP3 data. 
 
%\begin{figure}[h!t!b!]
%  \begin{center}
%    \includegraphics[width=70mm]{figures/results/low_accuracy_results_all2.ps}
%  \end{center}
%  \caption{Marginalized posteriors for both $\epsilon$ and $\xi$ free.}
%  \label{wmap1}
%\end{figure}

%\begin{figure}[h!t!b!]
%  \begin{center}
%    \includegraphics[width=40mm]{figures/epsilon1.ps}
%  \end{center}
%  \caption{Marginalized posterior of $\epsilon$ over a larger interval. Notice that it is not possible to give any upper constraint on this parameter. 
%The likelihood function of $\epsilon$ is much more complicated than that of $\xi$.}
%  \label{wmapepsilon}
%\end{figure}

\begin{figure}[h!t!b!]
  \begin{center}
    \includegraphics[width=75mm]{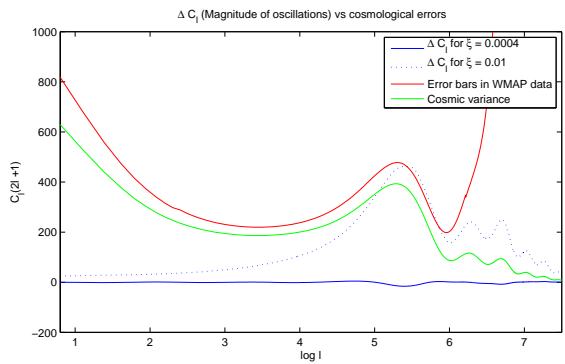}
  \end{center}
  \caption{The blue line shows the oscillations $\Delta C_\ell$ in the primordial power spectrum for $\xi = 0.0004$, $\gamma=0.01$ and $\epsilon=0.01$. The red bars are the error bars in the WMAP data, while the green graph is the cosmic variance. 
Notice how the modulations in the low-$\xi$ power spectrum (blue line) are much smaller than either. 
The dotted blue line represents a larger $\xi \sim 0.01$, which should be possible to detect, and gives upper constraints on
$\xi$.}
\label{figerror}
\end{figure}

For physically reasonable values of $\xi$ and $\epsilon$ ($\ll 1$), the amplitude of the modulations are well below $1\%$ of the total power spectrum.
Also, the cosmic variance and the errors in the WMAP data are several orders of magnitude greater than the amplitude of the modulations in
the power spectrum, as can be seen from figure \ref{figerror}. This explains why any detection of transplanckian effects in the WMAP data fails 
for low values of $\xi$.  

\section{Results from ideal CMB data }
\begin{figure}[h!t!b!]
  \begin{center}
    \includegraphics[width=30mm, height=20mm]{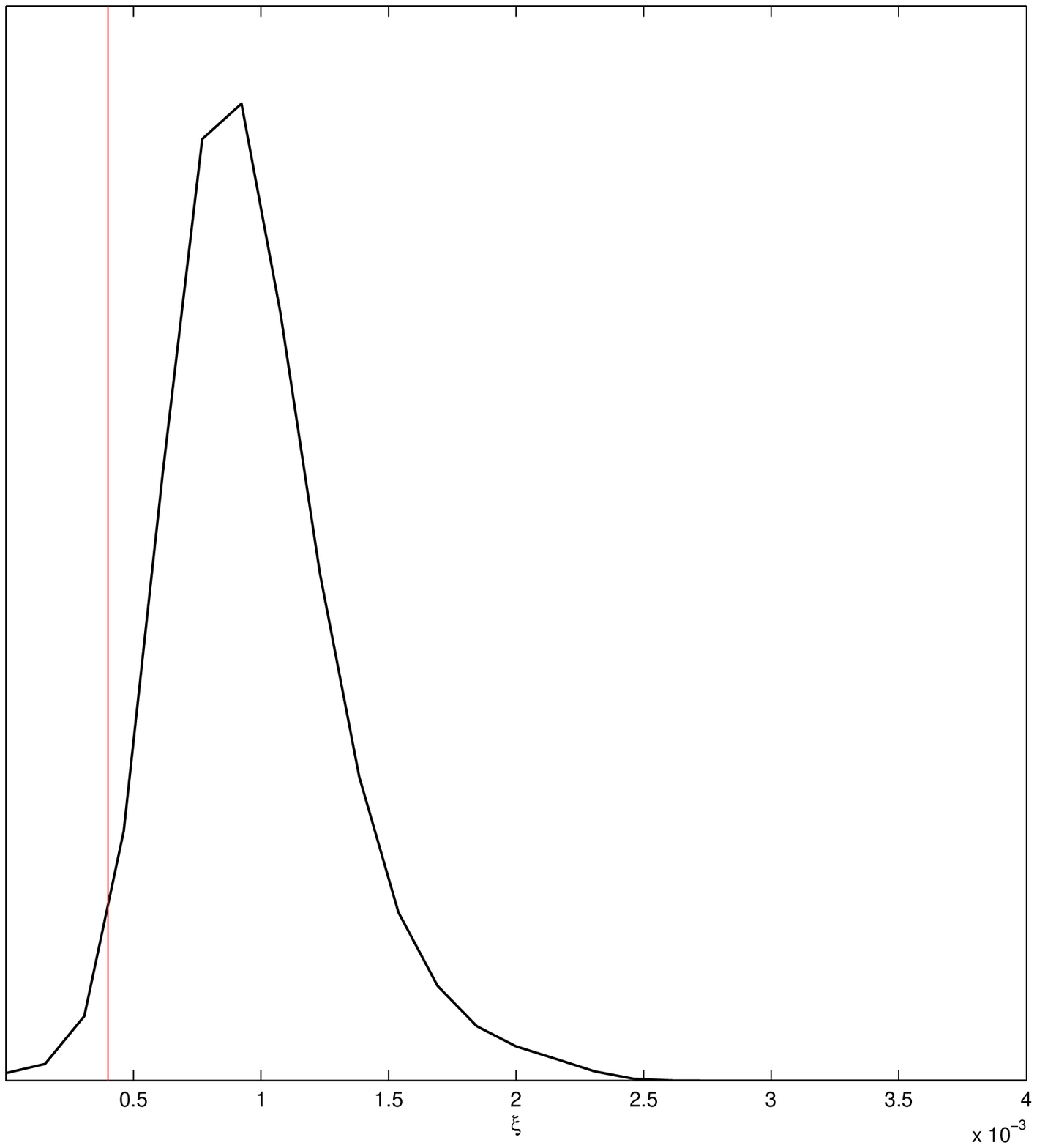}
    \includegraphics[width=30mm, height=20mm]{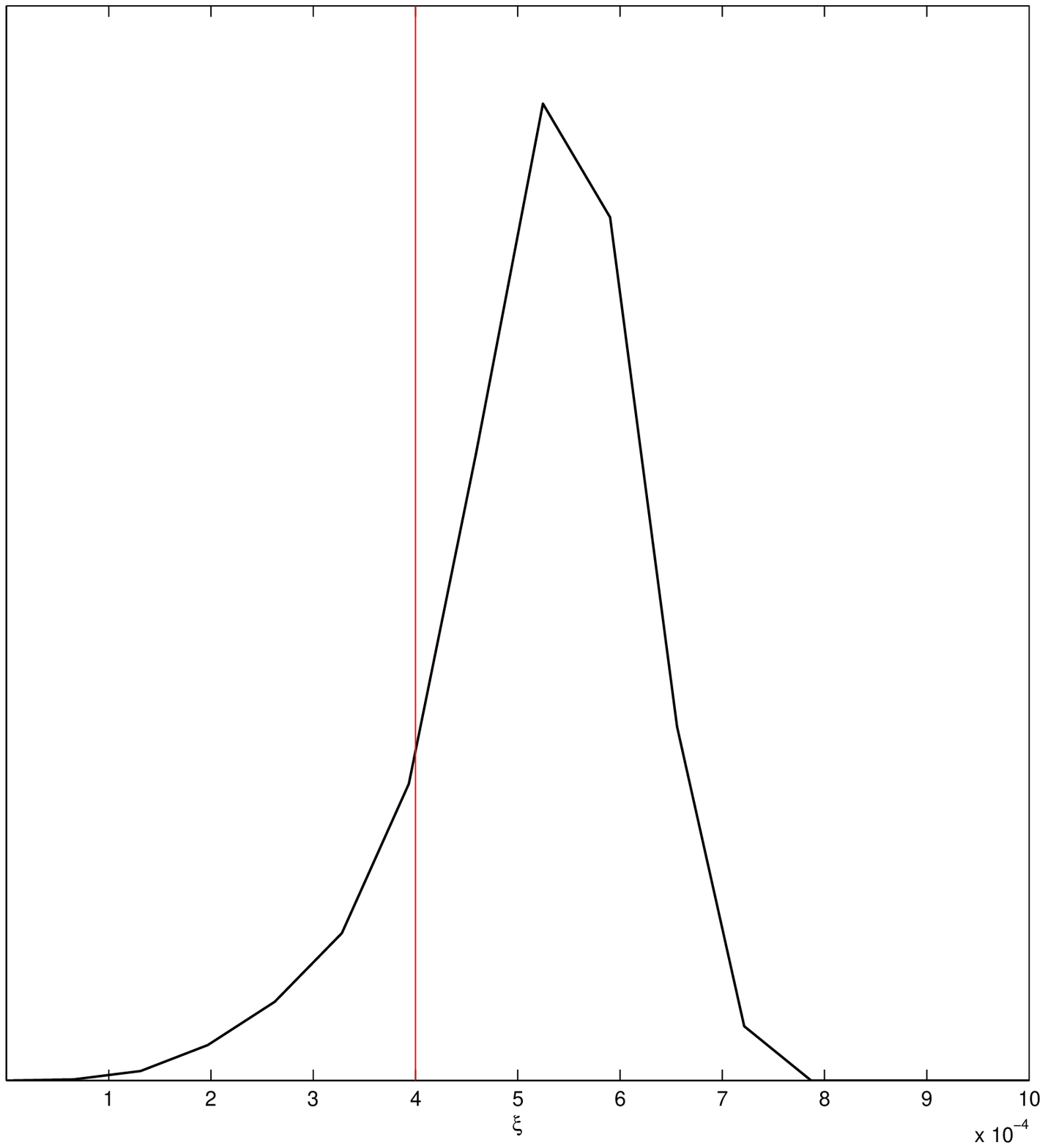}\\
    \includegraphics[width=30mm, height=20mm]{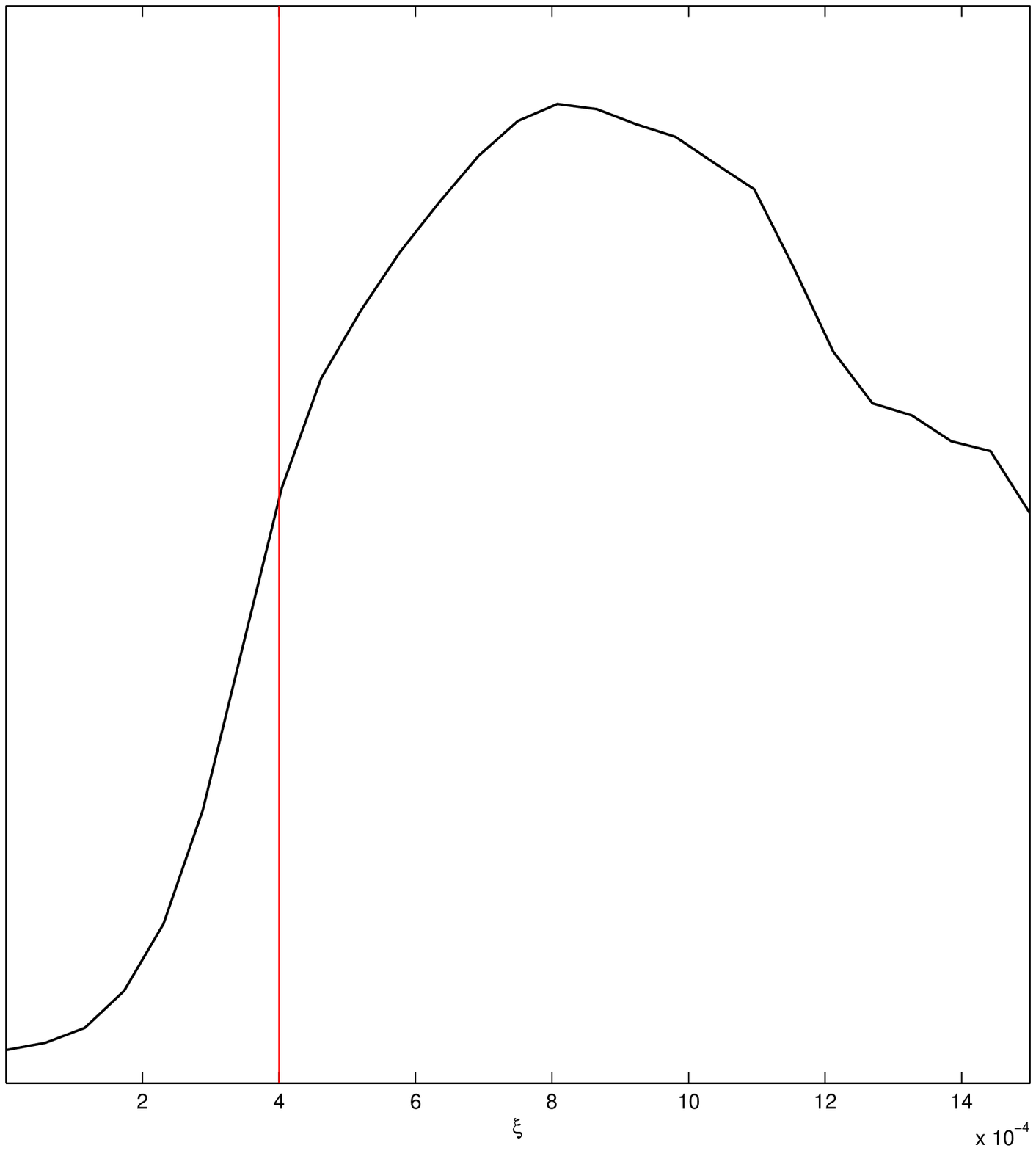}
    \includegraphics[width=30mm, height=20mm]{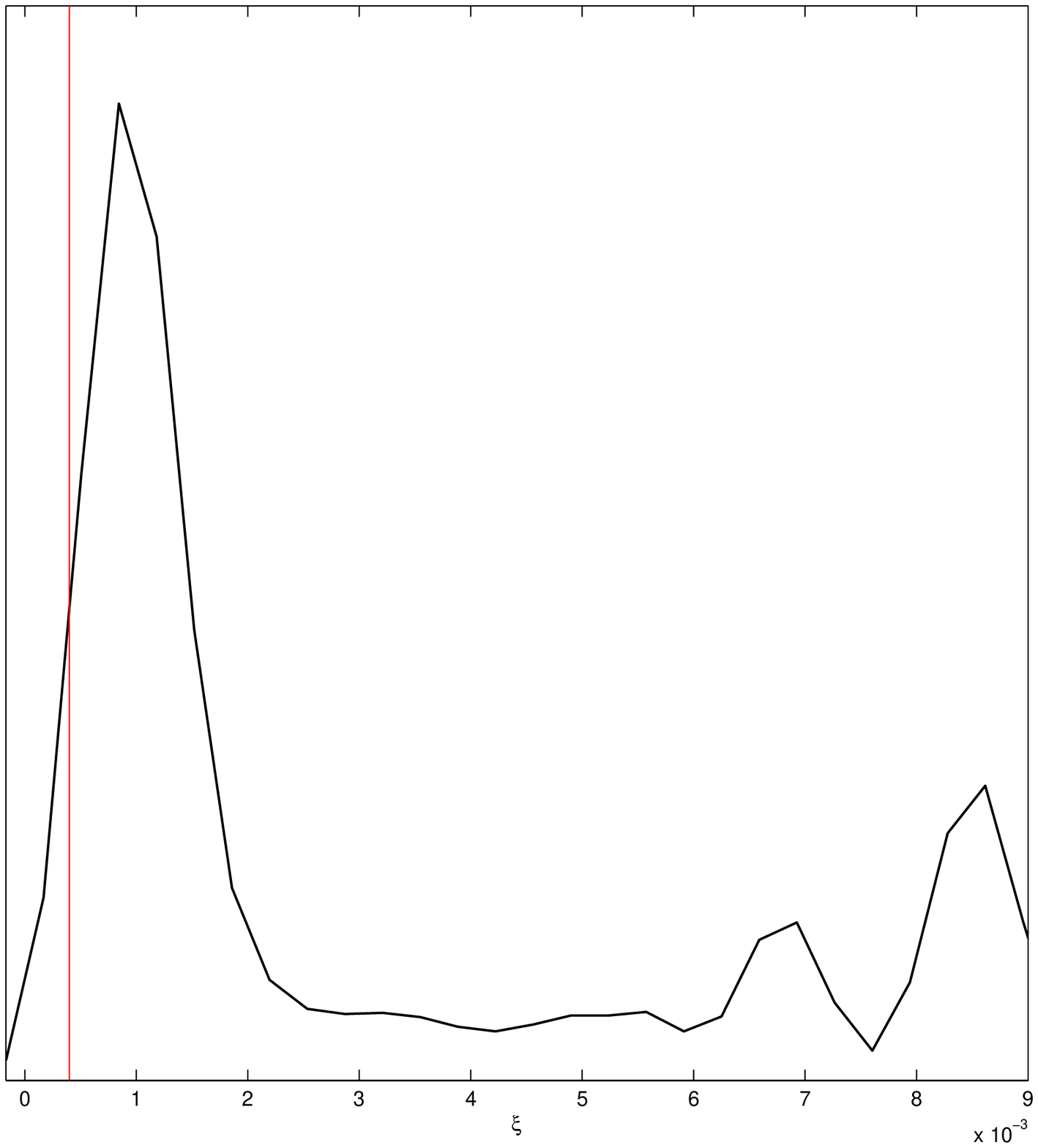}
  \end{center}
  \caption{
Top left: Marginalized posterior for $\xi$ with mean value at  $0.978 \cdot 10^{-3}$.
Top right: Marginalized posterior for $\xi$ with mean value at  $0.515 \cdot 10^{-3}$. 
Bottom left: Marginalized posterior for $\xi$ with no peak, almost uniform distribution.
Bottom right: Marginalized posterior for $\xi$ with several peaks. 
Notice how the peaks don't match, where the red line represents the Horava-Witten input
  parameter $\xi= 0.0004$ for the simulated data. }
 \label{figresultsxi}
\end{figure}
The WMAP3 data in CosmoMC was replaced by the optimal data set created from the power spectrum based on the Horava-Witten model.
The ideal $TT$ data are shown in figure \ref{simualteddata}. 

\subsection{Simulated TT data}
\begin{figure}[h!t!b!]
  \begin{center}
    \includegraphics[width=30mm, height=20mm]{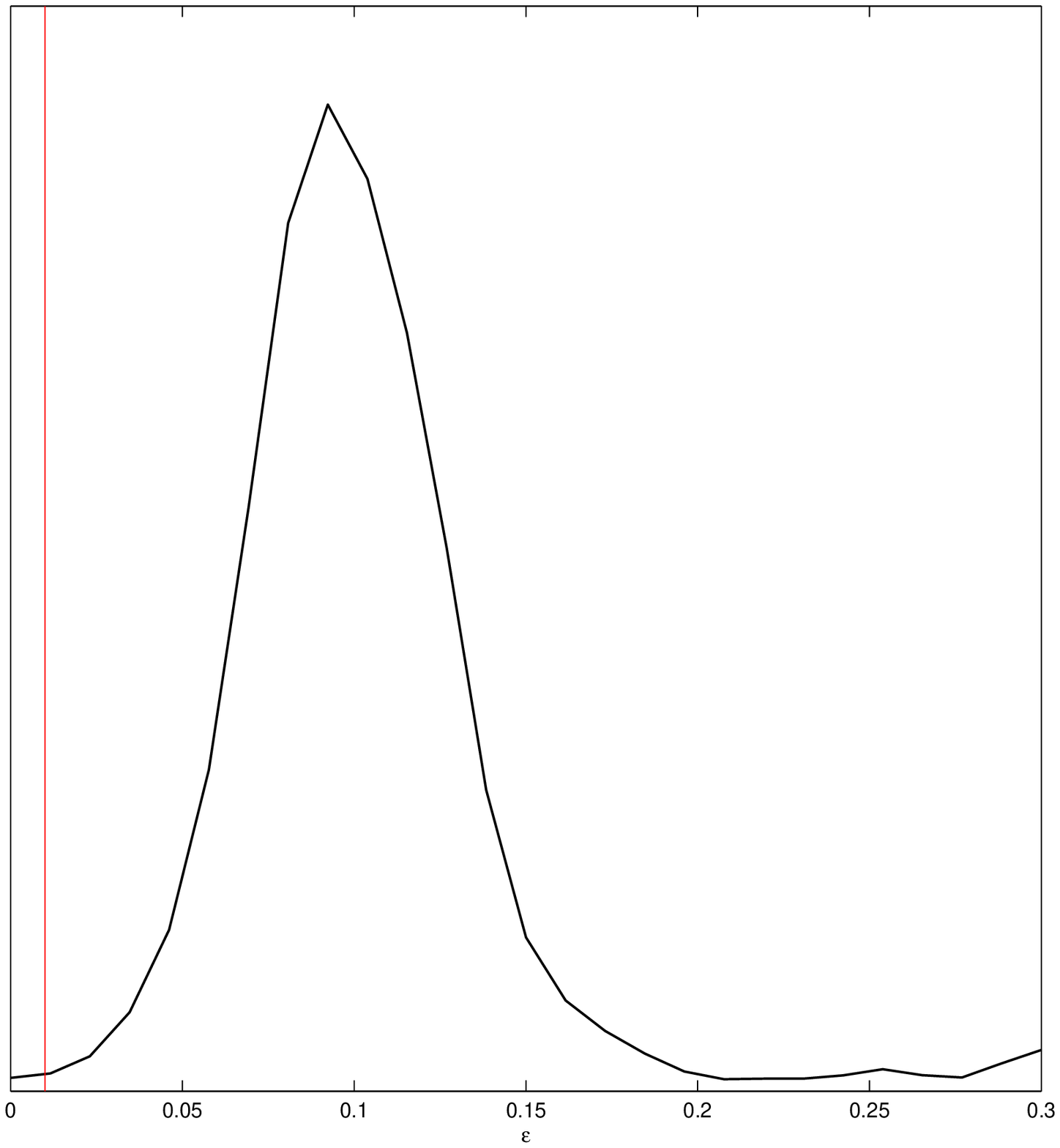}
    \includegraphics[width=30mm, height=20mm]{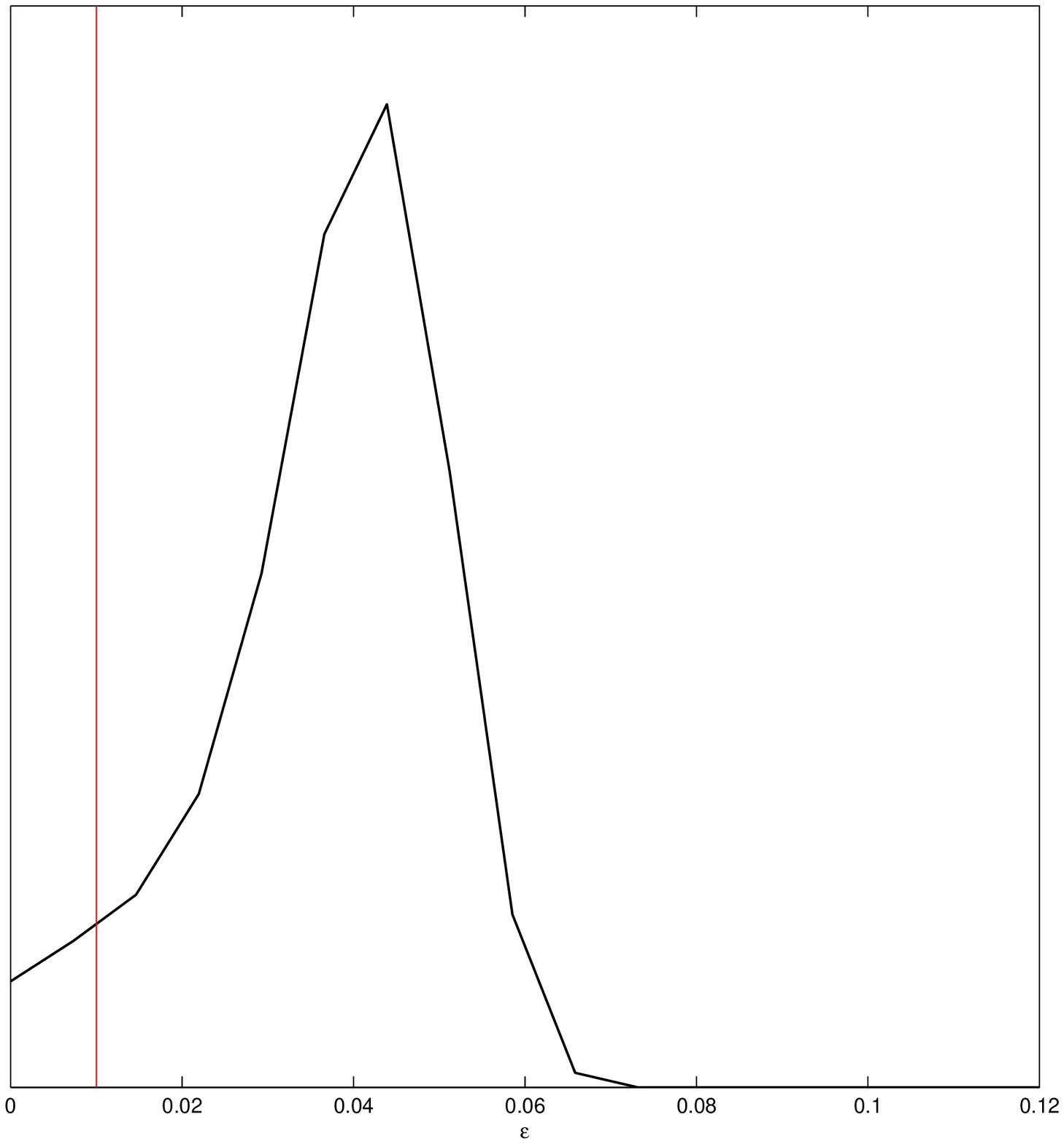}\\
    \includegraphics[width=30mm, height=20mm]{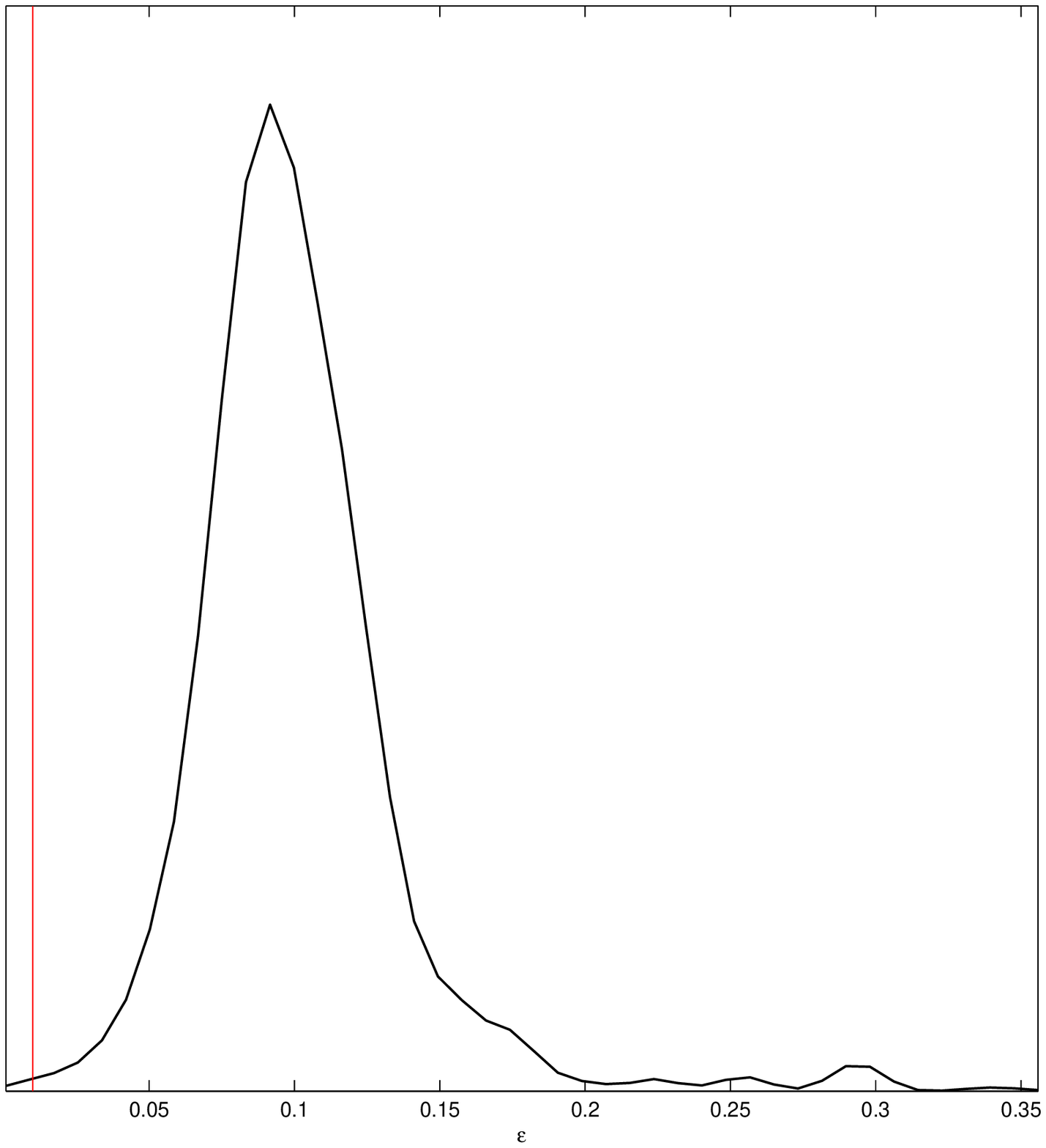}
  \end{center}
  \caption{
Top left: Marginalized posterior for $\epsilon$ with mean value at  $0.12$.
Top right: Marginalized posterior for $\epsilon$ with mean value at  $0.037$. 
Bottom left: Marginalized posterior for $\epsilon$ with mean peak at  $0.101$.
Notice how the peaks do not agree for different step lengths, where the red line represents the Horava-Witten input
  parameter $\epsilon= 0.01$ for the simulated data. 
}
 \label{figresultsepsilon}
\end{figure}

When performing parameter estimation of the transplanckian parameters $\xi$ and $\epsilon$ using the $TT$-data setup, 
the Monte Carlo Markov chain-method failed to converge 
to a stationary distribution. The resulting likelihoods for various $\xi$ and $\epsilon$ are presented in figure \ref{figresultsxi} and \ref{figresultsepsilon}, respectively. 
Notice
how different initial conditions and varying step length affect the end result: different initial parameters for the
same parameter results in different distributions. This suggests that the likelihood function which
decides the path of the random walkers traversing the cosmological parameter space has some troublesome properties.   

\subsection{Simulated TT,TE,EE and BB data}
\begin{figure}[h!t!b!]
  \begin{center}
    \includegraphics[width=70mm, height=55mm]{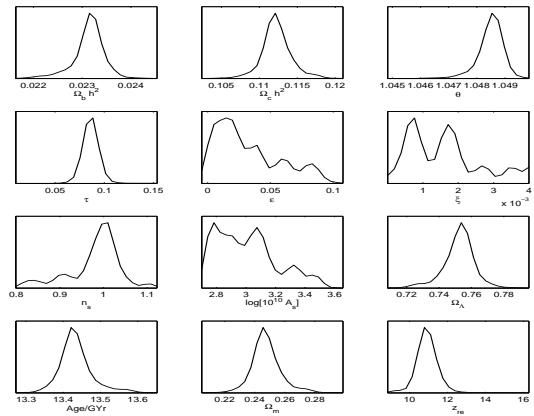}
  \end{center}
  \caption{Marginalized posteriors for the full TT, TE, EE and BB likelihood analysis using equation (\ref{likeli2})}
 \label{figfull}
\end{figure}

The marginalized posteriors can be seen in figure \ref{figfull}. Notice how $\tau$ is restored due to the inclusion of simulated TE and EE data. The transplanckian
parameters now reflect the ruggedness of the likelihood-function, and traces can also be seen in $n_s$ and $A_s$. As in the TT-case, the original
transplanckian input parameters are still not possible to pin down.

\begin{figure}[h!t!b!]
  \begin{center}
    \includegraphics[width=70mm, height=55mm]{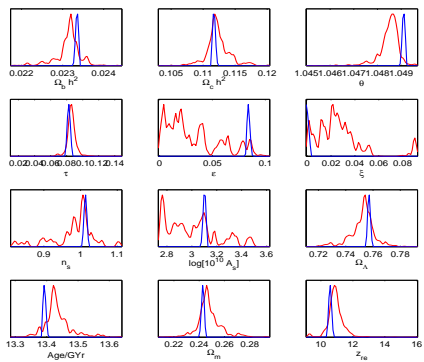}
  \end{center}
  \caption{Marginalized posteriors (red lines) with smaller bins than in figure \ref{figfull} for the full TT, TE, EE and BB likelihood analysis using equation (\ref{likeli2}). The likelihoods
are depicted with blue lines. Notice how the posteriors completely failed to converge to a distribution.}
 \label{figfull2}
\end{figure}
Figure \ref{figfull} and \ref{figfull2} show how the MCMC process fails in converging, as the marginalized
posterior show random walkers ``trapped'' in local minima.

\section{The exact likelihood function}
\begin{figure}[h!t!b!]
  \begin{center}
    \includegraphics[width=60mm]{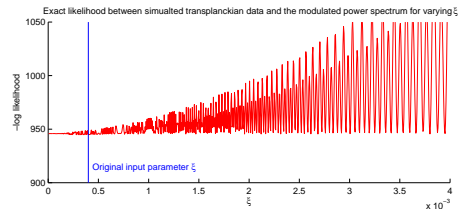}
  \end{center}
  \caption{$- \log \mathcal L$ as a function of $\xi$ for simulated transplanckian data.}
\label{likexi}
\end{figure}

\begin{figure}[h!t!b!]
  \begin{center}
    \includegraphics[width=60mm]{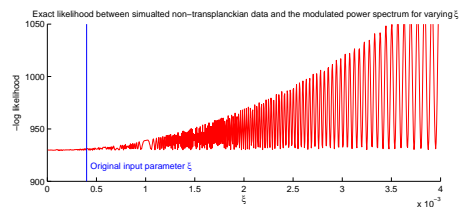}
  \end{center}
  \caption{$- \log \mathcal L$ as a function of $\xi$ for simulated non-transplanckian data. Notice how the non-transplanckian data is preferred
to the simulated transplanckian data set.}
\label{likexi2}
\end{figure}

\begin{figure}[h!t!b!]
  \begin{center}
    \includegraphics[width=60mm]{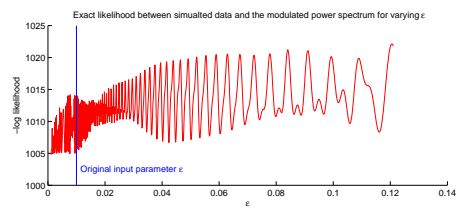}
  \end{center}
  \caption{$- \log \mathcal L$ between varying $\epsilon$ and simulated data.}
\label{likeepsilon}
\end{figure}

\begin{figure}[h!t!b!]
\begin{center}
    \includegraphics[width=60mm]{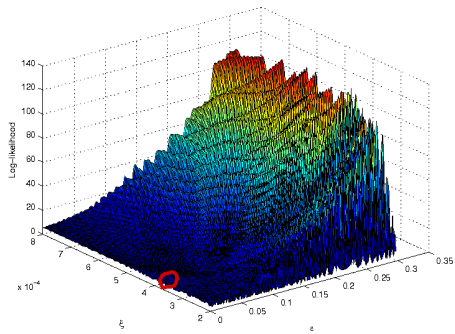}
  \end{center}
  \caption{$ -\log \mathcal L$ for varying both $\xi$ and $\epsilon$. The red ring represents the original input parameters $\xi =0.0004, \epsilon=0.01$ 
  for the simulated data set. Notice how a random walker would be trapped in any of the other local minima.}
\label{like2d}
\end{figure}

In order to explain why the parameter estimation fails, we investigated the likelihood function in the case of noiseless data 
and different power spectra for fixed cosmological parameters except the transplanckian $\xi$ and $\epsilon$. This
was done by grid integration, and no MCMC-sampling was involved. The resulting likelihood surfaces for continuous variations in $\xi$ and $\epsilon$ are seen
 in figure \ref{likexi} and \ref{likeepsilon}, respectively.  
A figure of the exact two-dimensional likelihood surface for both $\xi$ and $\epsilon$ 
is presented in figure \ref{like2d}.  
Figures \ref{likexi}, \ref{likexi2}, \ref{likeepsilon} and \ref{like2d} are $-\log \mathcal L$-plots, so the best-fit value corresponds with the global minimum.  
The likelihood surfaces are riddled with local minima, effectively rendering the usual MCMC-method useless: the
random walkers become ``trapped'' in the local holes, and will have problems converging to the correct stationary distribution. Take especially note of
the two-dimensional parameter space presented in figure \ref{like2d}, a landscape reminiscent that of an egg container.

The shape of the exact likelihood is a result of the \textit{nature} of the oscillations: equation (\ref{finalmp}) gives rise to a particularly nasty 
behaviour of the oscillations \footnote{Two
movies describing how continuous variations in $\epsilon$ and $\xi$ affect the superimposed oscillations can be downloaded from 
\textit{http://www.irio.co.uk/projects/thesis}}.
The problem with the likelihood function is not easily solved. In principle, one could perform a smoothing of the likelihood, something
which works well in a low-dimensional parameter space, but is hopelessly inefficient for a realistic number of parameters. \\

\subsection{Decoupling the amplitude}
\begin{figure}[h!t!b!]
\begin{center}
    \includegraphics[width=70mm]{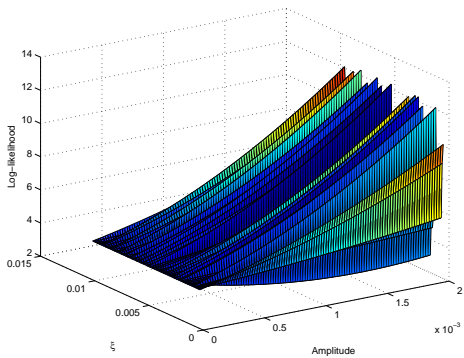}
  \end{center}
  \caption{-log Likelihood for varying $\xi$ and amplitude}
\label{figknoll}
\end{figure}
\begin{figure}[h!t!b!]
\begin{center}
    \includegraphics[width=70mm]{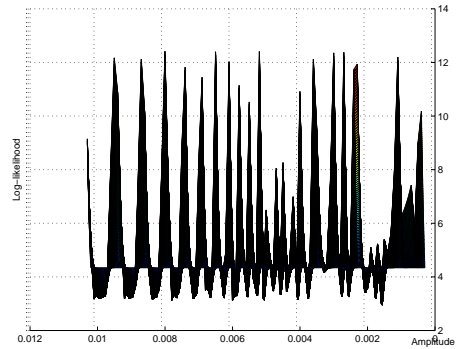}
  \end{center}
  \caption{-log Likelihood for varying $\xi$ and amplitude (different viewing angle)}
\label{figtott}
\end{figure}

The amplitude and the frequency of the transplanckian oscillations are linked in the model we have considered so far.  However, it is certainly possible that they might vary independently, and 
in \cite{jerome1, jerome2, jerome3, jerome4}, the authors operate with the amplitude of the transplanckian oscillations as a free amplitude. 
We repeated our analysis with frequency and amplitude decoupled.  
From
figure \ref{figknoll} and \ref{figtott} it should be clear that freeing the amplitude does not really add any significance to the likelihood
surface: the likelihood still gives the correct preferred amplitude.

\section{Bayesian evidence}

From the logarithmic likelihood in figure \ref{likexi} and \ref{likeepsilon}, it is obvious that it 
is impossible extract the values of the parameters of our model from the 
simulated data.  However, one might still ask the question of whether the 
data prefer the transplanckian model to the standard power-law primordial 
power spectrum.  In a Bayesian framework this question is answered by calculating the evidence ratio of the two models, see \cite{nestedsampling} for a 
nice introduction.  Here we fix $\epsilon = 0.01$ and use the evidence 
ratio to see whether the data prefer the introduction of the additional 
parameter $\xi$.  
We assume that the transplanckian parameter $\xi$ is \textit{not} degenerate with any of the cosmological parameters, designated by the parameter 
vector $\theta$. 
This might seem like a strong assumption, but we have checked that it is 
reasonable for the model and the parameters we consider in this paper. Also, 
the effects of primordial oscillations have been investigated in \cite{hamann-2007}, concluding that degeneracies 
of oscillations with standard cosmological parameters are virtually non-existent.

Let $\theta$ denote the non-transplanckian cosmological parameters, and $\hat \theta$
the best-fit $\Lambda CDM$ parameters. The logarithmic evidence is given by
\begin{equation}
  \Delta \ln E = \ln E - \ln E^0
\end{equation}
where $E^0$ is the evidence of the non-transplanckian model. Then
\begin{equation}
  \Delta \ln E = 
\langle \mathcal L \rangle_{\pi(\theta, \xi)} - 
\langle \mathcal L \rangle_{\pi(\theta)}
\end{equation}
\begin{equation}
  = 
\langle \mathcal L(\theta, \xi) \rangle_{\pi(\xi)} - 
\mathcal L (\hat \theta)
\end{equation}
where $\pi(\theta, \xi)$ is the prior over $(\theta, \xi)$ and $\pi(\theta)$ is the prior over $\theta$. Thus, $\langle \mathcal L \rangle_X$ denotes
the mean likelihood over the prior space $X$.
\begin{figure}[h!t!b!]
  \begin{center}
    \includegraphics[width=60mm]{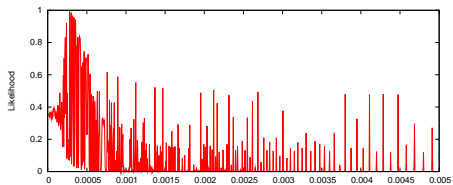}
  \end{center}
  \caption{Likelihood $\mathcal L(\xi)$ used in calculating the significant evidence.}
 \label{figlikelihoodjess}
\end{figure}
The likelihood can be found in figure \ref{figlikelihoodjess}, which is the exponential of the negative values in figure \ref{likexi}.
We choose the prior to be $\xi \in [0, 0.005]$. This results in
$\Delta \ln E \approx -1.5$ which shows that the transplanckian model is slightly \textbf{not} preferred. Hence even perfect CMB-measurements
over the whole sky does not give any significant evidence for transplanckian effects. When constraining the prior 
to reside in $\xi \in [0,0.0005]$, then $\Delta \ln E \approx 0.15$ and the transplanckian model is more or less
equally preferable to the non-transplanckian model. This is hardly surprising, as a continuous constraint of the prior 
$Pr(\xi) = \lim_{\xi \to 0} [0,\xi]$ results in $\Delta \ln E = \ln E^0 - \ln E^0 = 0$, which corresponds with the standard non-transplanckian model. 
The evidence never grows larger than $\sim 0.15$ for any prior constraint on $\xi$.

\section{Alternate multilevel sampling}
There exists a few multilevel sampling methods besides the standard MCMC/Metropolis-Hastings algorithm that can tame the ruffled likelihood through a series of 
optimizations. Reference \cite{montecarlo} contains a detailed introduction to these methods, most notably simulated annealing, umbrella sampling
and the simulated tempering method. 
%All these methods utilize an auxiliary ``temperature'' parameter that helps bridging parts of the likelihood surface where random walkers normally would become trapped in
%local hyper-surface minima. This way the likelihood surface becomes ``smoothed'', enabling the random walkers to reach any parts of the now accessible surface.
%However, in \cite{elgen}, a code based simulated annealing was used without much success, emphasizing the difficulties with the numerical implementations.  
%An alternative implementation of the simulated annealing technique might have worked. Investigating these alternative methods is a major undertaking which 
We proceed by a rough discussion on how some of these alternate sampling methods work when investigating the 1-dimensional likelihood presented in
figure \ref{likexi}.
\subsection{Metropolis-Hastings method, 1D}
\begin{figure}[h!t!b!]
  \begin{center}
    \includegraphics[width=60mm]{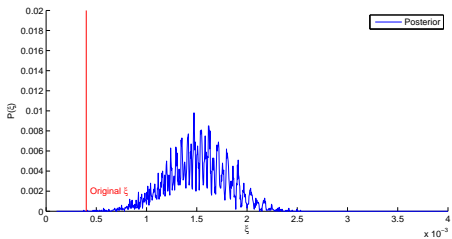}
  \end{center}
  \caption{The posterior of 10 000 MCMC walkers evaluated after 22 000 steps each in 1 dimension, with initial conditions $\xi_0 =0.0015$. None
  of the walkers reached the correct value of $\xi=0.0004$.}
 \label{figpost1}
\end{figure}

Figure \ref{figpost1} shows the posterior of 10 000 walkers after $\sim$ 22 000 steps each, using the Metropolis-Hastings algorithm. 
The likelihood is shown in figure \ref{likexi}. In theory,
the walkers should converge to a stationary distribution with the tallest peak located at $\xi= 0.0004$, but due to the ruffled
likelihood this will take forever. The initial values for the walkers was here  $\xi_0 = 0.0015$. This reflects what we have seen in figure 
 \ref{figresultsxi}, \ref{figresultsepsilon}, \ref{figfull} and \ref{figfull2} : that the complicated likelihood surface 
prevents the posterior to converge to the correct shape.

\subsection{Modified Metropolis-Hastings method, 1D}
\begin{figure}[h!t!b!]
  \begin{center}
    \includegraphics[width=60mm]{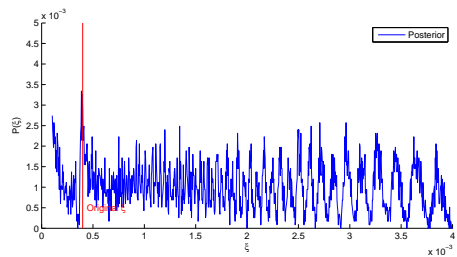}
  \end{center}
  \caption{The posterior of 10 000 MCMC walkers evaluated after 22 000 steps each in 1 dimension, with initial conditions $\xi_0 =0.0015$. The
posterior shows a clear contribution around the correct value of $\xi = 0.0004$.}
 \label{figpost2}
\end{figure}

Figure \ref{figpost2} shows the posterior of 10 000 walkers after $\sim $20 000 steps each, using the Metropolis-Hastings
algorithm. In addition, for each step there is a $0.01\%$ chance for a given walker to perform a ``large'' jump to a new, (uniformly) random location.
The walkers will therefore tend to accumulate at the correct location (where the -log likelihood is lowest, $\xi=0.0004$). This is a slow
method, and needs a fair amount of steps to give results. It is also not especially ``robust'', as the accumulation of walkers at the 
lowest  point is dependent on fine-tuning the large-step rate (here, $0.01\%$).

\subsection{Simulated annealing, 1D}
\begin{figure}[h!t!b!]
  \begin{center}
    \includegraphics[width=60mm]{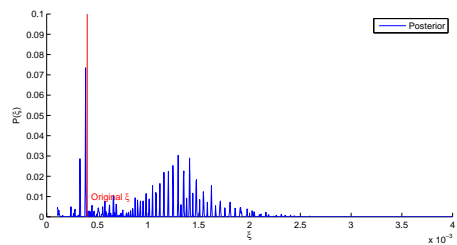}
  \end{center}
  \caption{The result of 10 000 MCMC walkers evaluated after 6 000 steps in 1 dimension, with initial conditions $\xi_0 =0.0015$ (not a posterior). The temperature
was slowly decreased until reaching $T=0$, where the correct value $\xi = 0.0004$ was obtained. }
 \label{figpost3}
\end{figure}

A description of the simulated annealing (SA) method can be found in \cite{montecarlo}. As in the previous examples, 10 000
random walkers were initialized at $\xi_0 = 0.0015$. The distribution spread out evenly as the SA temperature $T$ was high, and
started converging to different minima as $T$ was lowered. As $T \to 0$, a good portion of walkers ended up
at $\xi = 0.0004$, as seein in figure \ref{figpost3}. This shows that the SA method can be successfully used to determine the correct minima for this type
of likelihood landscape.  \\

However, the implementation and investigating of these alternative methods in CosmoMC is a major undertaking which we hope to pursue in future work.

\section{Conclusion}
We have shown that standard MCMC methods are not suitable for detecting transplanckian oscillations in the CMB. 
The 3-year WMAP3 data alone gives weak upper constraints on the transplanckian parameters. With the simulated noiseless data, 
the results are more ambiguous. In theory, the estimation of the transplanckian parameters \textit{should} be possible, but the
messy likelihood function renders conventional MCMC methods of little use. 
This is not due to the amplitude of the superimposed oscillations in the power spectrum alone, but
rather reflects the chaotic behaviour of the oscillations for varying transplanckian parameters $\xi$ and $\epsilon$. Small variations in the parameter space thus
correspond to large variations in
the likelihood function, as presented in figure \ref{likexi} and \ref{likeepsilon}. As in \cite{elgen}, we conclude that 
CMB data are highly sensitive to these modulations in the primordial power spectrum, but it is virtually impossible to perform
a parameter estimation from WMAP3 data or high-precision simulated data using conventional MCMC methods. We have also
shown, via Bayesian analysis, that even a high-precision measurement of the full CMB sky does not give any significant evidence
for transplanckian effects. It should however be possible to implement the SA-method in CosmoMC when using simulated high-precision data
in order to detect the transplanckian parameters.

We have in this paper shown that it is very unlikely that CMB-data could support the establishment of
transplanckian effects, even for models where quantum gravity is relevant 
on relatively low energy scales, like in the Horava-Witten model. 
There are several reasons for this lamentable conclusion:
\begin{itemize}
 \item A low quantum gravity scale, for example the value  $\xi = 0.0004$ chosen in our simulations, gives a very low amplitude of the oscillations in the primordial power spectrum.
   Hence, cosmic variance diffuses the (assumed noiseless) signal.
 \item Evidence can maximally rise above $0.15$, showing that the transplanckian model is only very weakly preferred when assuming a 
   Horava-Witten universe. 
 \item Even large (and unphysical) values of $\xi$ will result in good fits for singular points due to the behaviour of the oscillations (they
   periodically reach zero when $\sin x = 0$)\\
 \item The conventional MCMC-methods in CosmoMC fails, and we have no current computational methods implemented for solving the problem of transversing
   the extremely complicated likelihood surface. 
\end{itemize}
   
\acknowledgments

{\O}E acknowledges support from the Research Council of Norway, project number 162830. 
We thank H. K. Eriksen and J. R. Kristiansen for valuable discussions.

%%%%%%%%%%%%%%%%%%%%%%%%%%%%%%%%%

%\bibliography{cites}

\end{document}